\documentclass[a4paper]{amsart}
\usepackage[utf8]{inputenc}
\usepackage{tikz,eucal,hyperref}
\usetikzlibrary{matrix,arrows,shapes,calc,intersections,decorations.markings,decorations.pathmorphing}

\usepackage{tikz-cd}
\usepackage{tkz-euclide}
\usepackage{amssymb,mathtools}

\tikzset{->-/.style={decoration={
  markings,
  mark=at position #1 with {\arrow{stealth'}}},postaction={decorate}}}
  
\tikzset{proj/.style={circle,draw=black, fill=white,inner sep=0pt}}
\tikzset{anch/.style={circle,fill=black,draw=black,inner sep=0pt,minimum size=1mm}}
\tikzset{pnt/.style={circle,fill=black,draw=white,inner sep=1pt}}

\theoremstyle{plain}
  \newtheorem{thm}{Theorem}[section]

  \newtheorem*{defn*}{Definition}

  \newtheorem{lem}[thm]{Lemma}
\theoremstyle{definition}
  
  \newtheorem*{rem}{Remark}

\newcommand{\mf}{\mathfrak}
\newcommand{\mc}{\mathcal}

\newcommand{\on}{\operatorname}
\newcommand{\g}{\mathfrak{g}}
\newcommand{\h}{\mathfrak{h}}
\newcommand{\bw}{{\textstyle\bigwedge}}

\newcommand{\la}{\langle}
\newcommand{\ra}{\rangle}
\newcommand{\R}{\mathbb{R}}

\newcommand{\C}{\mathbb{C}}
\newcommand{\CP}{\mathbb{CP}^1}

\newcommand{\id}{\on{id}}
\newcommand{\cf}{\mathsf{Conf}}

\title[RG flow of Chern-Simons boundary conditions]{Renormalization group flow of Chern-Simons boundary conditions and generalized Ricci tensor}
\thanks{Supported in part by the  NCCR SwissMAP of the Swiss National Science Foundation.}
\author{Ján Pulmann}
\address{Section of Mathematics, University of Geneva, Switzerland}
\email{jan.pulmann@unige.ch}
\author{Pavol \v{S}evera}
\address{Section of Mathematics, University of Geneva, Switzerland}
\email{pavol.severa@gmail.com}
\author{Donald R. Youmans}
\address{Section of Mathematics, University of Geneva, Switzerland}
\email{donald.youmans@unige.ch}
\begin{document}

\maketitle

\begin{abstract}
We find a Chern-Simons propagator on the ball with the chiral boundary condition. We use it to study perturbatively Chern-Simons boundary conditions related to 2-dim $\sigma$-models and to Poisson-Lie T-duality. In particular, we find their renormalization group flow, given by the generalized Ricci tensor. Finally we briefly discuss what happens when the Chern-Simons theory is replaced by a Courant $\sigma$-model or possibly by a more general AKSZ model.
\end{abstract}

\section{Introduction}
The chiral boundary condition, introduced by Witten in his famous work \cite{W}, is arguably the most useful boundary condition of Chern-Simons theory. Roughly speaking, it needs the boundary $\Sigma$ of the 3-manifold $M$ to be a Riemann surface and it requires the connection $A\in\Omega^1(M,\g)$ to be in
$\Omega^{1,0}(\Sigma,\g)$
when restricted to $\Sigma$. The boundary field theory thus obtained is the chiral WZW model.

There is a straightforward generalization of the chiral boundary condition, introduced in \cite{S}, linking Chern-Simons theory to other 2-dim $\sigma$-models and to Poisson-Lie T-duality. Given an orthogonal splitting $\g=V_+\oplus V_-$ to a pair of vector subspaces, such that the pairing on $\g$ is positive definite on $V_+$ and negative definite on $V_-$, the corresponding \emph{$V_+$-boundary condition} requires the restriction of $A$ to $\Sigma$ to be in $$\Omega^{0,1}(\Sigma,V_+)\oplus\Omega^{1,0}(\Sigma,V_-).$$

To get a 2-dim $\sigma$-model we then need to consider the Chern-Simons theory on $\Sigma\times[0,1]$, with the $V_+$-boundary condition on $\Sigma\times\{0\}$, and with a topological boundary condition on $\Sigma\times\{1\}$ given by a Lagrangian Lie subalgebra $\h\subset\g$. 
$$
\begin{tikzpicture}
\coordinate (sh) at (0,-1.3);
\fill[red!10!white] (0,0)--(5,0)--(6,1)--(1,1)--cycle;
\fill[yshift=-1.3cm,blue!10!white] (0,0)--(5,0)--(6,1)--(1,1)--cycle;
\draw[yshift=-1.3cm, gray, very thin](6,1)--(1,1)--(0,0);
\draw[gray, very thin] (1,1)--+(sh);
\draw (0,0)--(5,0)--(6,1)--(1,1)--cycle;
\draw[yshift=-1.3cm] (0,0)--(5,0)--(6,1);
\draw[very thick,->,red!30!white](-0.5,0.5)node[left,black]{$\h$-b.c.}--(0.4,0.5);
\draw[very thick,->,blue!30!white,yshift=-1.3cm](-0.5,0.5)node[black,left]{$V_+$-b.c.}--(0.4,0.5);
\draw (0,0)--+(sh) (5,0)--+(sh) (6,1)--node[right]{$\Sigma\times [0,1]$}+(sh);

\begin{scope}[yshift=-2.5cm]
\draw[fill=black!5!white] (0,0)--(5,0)--(6,1)--(1,1)--cycle;
\draw (6,0.5)node{$\Sigma$};
\end{scope}
\end{tikzpicture}
$$
This ``Chern-Simons sandwich'' is equivalent to a 2-dim $\sigma$-model with the worldsheet $\Sigma$ and the target $G/H$. Different choices of $\h$  give different targets, all of them linked by Poisson-Lie T-duality (the usual T-duality corresponds to $G$ being a torus).

It is thus natural to study the $V_+$-boundary condition on its own, since that is where the dynamical degrees of freedom of the $\sigma$-models live; the $\sigma$-models see $\h$ only through topological degrees of freedom.

For a general $V_+$ the conformal invariance of the $V_+$-boundary condition gets broken by perturbative quantization, and we should get a renormalization group (RG) flow of the boundary condition, i.e.\ of the subspace $V_+\subset\g$. It should give us the RG flow (i.e.\ the Ricci flow) of the 2-dim $\sigma$-models equivalent to the Chern-Simons sandwiches; indeed, the RG flow should only affect the $V_+$-boundary condition, and not the bulk Chern-Simons theory or its topological boundary conditions.

In this work we study perturbatively the Chern-Simons theory on a ball with the $V_+$-boundary condition. We find a natural $SL_2(\C)$-invariant propagator \eqref{gmprop}, which is of independent interest, as it applies also to the chiral boundary condition. It turns out that only one 1-loop diagram is potentially divergent \eqref{eyediagram}, giving rise to the 1-loop Weyl anomaly \eqref{conf-anom} on the boundary sphere, and to the RG flow \eqref{RG} of $V_+$. This RG flow is given by the so called \emph{generalized Ricci tensor} \cite{CSW,G,JV,SV}, which thus gets a natural meaning via the Chern-Simons theory. The RG flow also agrees with the one found in \cite{SST} for a certain 2-dim model which can be seen, in some sense, as a Hamiltonian version of the $V_+$-boundary condition. As expected, it is compatible with the Ricci flow on the targets $G/H$, as shown already in \cite{SV}.

At the end of the paper we briefly discuss what happens when the Chern-Simons theory is replaced by a more general Courant $\sigma$-model; the RG flow of the boundary condition still turns out to be given by the generalized Ricci tensor \eqref{ggric} (which is the usual Ricci tensor in the case of exact Courant algebroids). We also briefly mention how to deal with the RG flow of boundary conditions of more general TFTs of the AKSZ type \cite{AKSZ} by computing the divergent part of the effective action \eqref{akszrg}.\bigskip

\noindent\textbf{Acknowledgements.} We are grateful to A.\ Cattaneo, K.\ Gawedzki, P.\ Safronov, and especially to P.\ Mnev, for stimulating discussions.

	
	

\section{Chern-Simons propagator with the chiral boundary condition}

In this section we shall construct a propagator for the Chern-Simons theory on the unit closed ball $B^3$, with the chiral boundary condition familiar from the CS/WZW correspondence \cite{W}. We shall also find a propagator with a slightly modified chiral boundary condition, which we call $V_+$-boundary condition.

The (naive) space of fields of the Chern-Simons theory is $\Omega^1(B^3,\g)$. The Lie algebra $\g$ will play a very little role in the propagator, so for the moment we shall suppose that the space of fields is $\Omega^1(B^3)$ and restore $\g$ only at the  end of this section.

\subsection{Chiral boundary condition}
The chiral boundary condition demands that the fields, when restricted to $\partial B^3 = S^2$, are in $\Omega^{1,0}(S^2)\subset\Omega^1(S^2)$ (using the standard complex structure on $S^2=\CP$). 

Let us rephrase this boundary condition in the Batalin-Vilkovisky (BV) formalism, including ghosts (0-forms) and antifields (2-forms and 3-forms). Let 
$$\Omega_b(B^3)\subset\Omega(B^3)$$
be the (acyclic) subcomplex of the forms whose restriction to $S^2$ is in
$$\Omega_{(-)}(S^2):=\Omega^{1,0}(S^2)\oplus\Omega^2(S^2)\ \subset\ \Omega(S^2).$$
Then our BV space of fields is
$\Omega_b(B^3)[1]$.


For future reference, let 
$$\Omega_{\bar b}(B^3)\subset\Omega(B^3)$$
be the subbcomplex given by the boundary condition
$$\Omega_{(+)}(S^2):=\Omega^{0,1}(S^2)\oplus\Omega^2(S^2)\ \subset\ \Omega(S^2).$$

\subsection{Propagator}\label{ssec:Propagator}
The propagator we are looking for should be a degree $-1$ map 
$$h\colon\Omega_b(B^3)\to\Omega_b(B^3)$$
such that
$$d\,h+h\,d=\id.$$
Moreover, $h$ should be given by an integral kernel 
$$P_0\in\Omega^2(\cf_2(B^3)),$$
where
$$\cf_2(B^3):=B^3\times B^3\setminus\text{diagonal}.$$

Let us describe a suitable $P_0$ explicitly, using the natural $SL(2,\C)$-invariant holomorphic closed 2-form
\begin{equation}\label{omega}
\omega =\frac{dz_1\wedge dz_2}{(z_1-z_2)^2}\ \in\ \Omega^2(\cf_2(\CP)). 
\end{equation}
Given any 2 points 
$$q_1\neq q_2\in B^3$$ let $\gamma$ be the hyperbolic geodesic passing through $q_1$ and $q_2$, where we view $B^3$ as the Poincaré model of the hyperbolic space. Let 
$$z_1,z_2\in\CP=S^2$$
be the points where $\gamma$ intersects $S^2$, $z_1$ on the side of $q_1$ and $z_2$ on the side of $q_2$. 
$$
\begin{tikzpicture}[scale=1.5]
  \tkzDefPoint(0,0){O}
  \tkzDefPoint(1,0){A}
  \tkzDrawCircle(O,A)  
  \tkzDefPoint(-.1,-.5){q1}
  \tkzDefPoint(.7,0.1){q2}
  \begin{scope}
    \tkzClipCircle(O,A)
    \tkzDefCircle[orthogonal through=q1 and q2](O,A) \tkzGetPoint{B}
    \tkzInterCC(O,A)(B,q1) \tkzGetPoints{C}{D}  
    \tkzDrawCircle[color=red, orthogonal through=q1 and q2](O,A)%
  \end{scope}
  \tkzDrawPoints[size=4](q1,q2,C,D) 
  \tkzLabelPoint[left](q1){$q_1$}
  \tkzLabelPoint[above](q2){$q_2$}
  \tkzLabelPoint[below](D){$z_1$}
  \tkzLabelPoint[right](C){$z_2$}
\end{tikzpicture}
$$
We thus obtain a map
$$r\colon\cf_2(B^3)\to \cf_2(\CP),\quad r(q_1,q_2):=(z_1,z_2)$$
and we set 
\begin{equation}\label{P0}
P_0:=r^*\omega/2\pi i.
\end{equation}

One can easily see that $P_0$ has the required properties. In particular, if we fix $q_1$  to be the center of $B^3$, $P_0$ becomes  the (normalized) area form on $S^2$ pulled back to $B^3\setminus\{\text{center}\}$ along the radii. By construction, $P_0$ is $SL(2,\C)$-invariant, where $SL(2,\C)$ acts on $B^3$ via the hyperbolic isometries.

\begin{rem}
In the usual setup of BV quantization one needs to choose a Lagrangian submanifold in the space of fields and restrict the path integral to this submanifold. The corresponding homotopy $h$ then satisfies $h^2=0$. We do not verify this relation for our $h$; as explained in \cite{CM}, it is not necessary.
\end{rem}

\subsection{The case of a non-trivial $\g$}
For a general Lie algebra $\g$ with a non-degenerate invariant symmetric pairing  $\la\,,\ra$ (which is negative-definite in the usual case of compact semisimple $\g$'s), the space of Chern-Simons fields with the chiral boundary condition is $\Omega_b(B^3,\g)[1]$, and the propagator is simply
$$P=P_0\otimes t\in \Omega^2\bigl(\cf_2(B^3),\g\otimes\g\bigr)$$
where $t\in\g\otimes\g$ is the inverse of the pairing $\la\,,\ra$. 
\subsection{Generalized metrics}\label{sec:gm}
Let us suppose that the Lie algebra $\g$ is split to an orthogonal sum
$$\g=V_+\oplus V_-,\quad V_-=V_+^\perp$$
of vector subspaces, with $\la\,,\ra|_{V_+}$ positive-definite and $\la\,,\ra|_{V_-}$ negative-definite. Such a splitting is called a \emph{generalized metric}  on $\g$ (see e.g.\ \cite{Gu}). We can use it to get  a minor generalization of the chiral boundary condition, which we shall call the \emph{$V_+$-boundary condition}: it is given by the subcomplex
$$\Omega_{(+)}(S^2,V_+) \oplus \Omega_{(-)}(S^2,V_-) \subset\Omega(S^2,\g),$$
i.e.\ the  space of fields with this boundary condition is
$$\Omega_{\bar b}(B^3,V_+)[1] \oplus \Omega_{ b}(B^3,V_-)[1]\subset \Omega(B^3,\g)[1].
$$

The corresponding propagator is
\begin{equation}\label{gmprop}
P=\bar P_0\otimes t_+ +  P_0\otimes t_-\in \Omega^2\bigl(\cf_2(B^3),\g\otimes\g\bigr)
\end{equation}
where $t_+\in V_+\otimes V_+$ and $t_-\in V_-\otimes V_-$ is the inverse of the pairing on $V_+$ and on $V_-$ respectively (we have $t = t_+ + t_-$).


\section{2-dim $\sigma$-models on Chern-Simons boundary and Poisson-Lie T-duality}\label{sec:PL}

In this section we shall explain the link between the Chern-Simons theory with the $V_+$-boundary condition, 2-dim $\sigma$-models, and Poisson-Lie T-duality, following \cite{PSV,S}.

\subsection{A tale of two boundary conditions}
If we want to study the Chern-Simons theory on a compact oriented 3-manifold $M$ with boundary $\Sigma$, we need to choose a boundary condition. In general, such a boundary condition is given by a Lagrangian differential graded submanifold in the space of the boundary fields $\Omega(\Sigma,\g)[1]$, and we need to add to the usual Chern-Simons action functional
\begin{equation}\label{CSaction}
S(A) = \int_M \frac12 \la A,dA \ra + \frac16 \la A, [A,A] \ra
\end{equation}
a boundary term whose variation is $-\frac12\int_\Sigma\la\delta A,A\ra$. 
In the following, however, we will be only interested in two special boundary conditions for which the  boundary terms  vanish and hence it suffices to consider the action \eqref{CSaction}.

\subsubsection{A (classically) topological boundary condition}
Let us choose a Lagrangian Lie subalgebra $\h\subset\g$ (i.e.\ $\h^\perp=\h$) and let us consider the boundary condition given by the subcomplex
\begin{equation}\label{h-bc}
\Omega(\Sigma,\h)\subset\Omega(\Sigma,\g).
\end{equation}
This boundary condition doesn't use any geometric structure on $\Sigma$, so it is (classically) topological. 
 We shall call it \emph{$\h$-boundary condition}.

The $\h$-boundary condition leaves us with a residual gauge symmetry on $\Sigma$, with gauge transformations taking values in $H$ (while in the bulk gauge transformations take values in $G$). The corresponding ghost fields are given by the 0-form part of the boundary condition, $\Omega^0(\Sigma,\h)$.

\begin{rem}
If $\h$ is not unimodular, the $\h$-boundary condition seems to have a 1-loop anomaly stemming from potential tadpole diagrams. Let us conjecture that we can still make sense of it by choosing an area form on $\Sigma$ up to a constant multiple, at the price that the boundary condition will no longer be topological (if true, the boundary condition would be ``rigid-scale invariant'', but not Weyl-invariant). 
\end{rem}

\subsubsection{Generalized metrics, again}\label{subsubsec:generalized_metrics_bc}
Let us now consider the $V_+$-boundary condition for an arbitrary compact oriented 3-manifold $M$ with boundary $\Sigma$. 
We need to choose a conformal structure on $\Sigma$. 
The $V_+$-boundary condition is then given by the subcomplex
$$\Omega_{(+)}(\Sigma,V_+) \oplus \Omega_{(-)}(\Sigma,V_-) \subset\Omega(\Sigma,\g)$$
where
\begin{align*}
\Omega_{(+)}(\Sigma)&:=\Omega^{0,1}(\Sigma)\oplus\Omega^2(\Sigma)\subset\Omega(\Sigma),\\
\Omega_{(-)}(\Sigma)&:=\Omega^{1,0}(\Sigma)\oplus\Omega^2(\Sigma)\subset\Omega(\Sigma).
\end{align*}


Unlike the $\h$-boundary condition, the $V_+$-boundary condition completely removes the gauge symmetry on the boundary. This is reflected in the absence of ghosts on the boundary. Classically, the $V_+$-boundary condition is conformally invariant, but for a general $V_+$ this invariance gets broken by perturbative quantization (see \S\ref{sec:diag}).

\subsection{2-dim $\sigma$-models on Chern-Simons boundary}

Let us now consider the 3-manifold 
$$M=\Sigma\times [0,1],$$
where $\Sigma$ is a Riemann surface.
The boundary of $M$ has 2 components, $\Sigma\times\{0\}$ and $\Sigma\times\{1\}$.
Let us impose the $V_+$-boundary condition on $\Sigma\times\{0\}$ and the $\h$-boundary condition on $\Sigma\times\{1\}$. The gauge transformations are thus trivial on $\Sigma\times\{0\}$ and take values in $H$ on $\Sigma\times\{1\}$. 

As explained in \cite{PSV,S}, Chern-Simons theory on $M=\Sigma\times[0,1]$, with this choice of boundary conditions, is equivalent to a 2-dim $\sigma$-model with worldsheet $\Sigma$ and target space $G/H$. Namely, given points $z_1,\dots,z_n\in\Sigma$ and functions $f_1,\dots,f_n\in C^\infty(G/H)$, which we can see as $H$-invariant functions on $G$, we have the following equality of correlation functions:
\begin{equation}
\bigl\la f_1(\phi(z_1))\dots f_n(\phi(z_n))\bigr\ra_{\sigma\text{-model}}= \bigl\la f_1(\on{hol}_{\gamma_1})\dots f_n(\on{hol}_{\gamma_n}) \bigr\ra_\text{Chern-Simons}.
\end{equation}
Here $\phi \colon \Sigma\to G/H$ is the field of the $\sigma$-model, $\gamma_i$ is the path $\{z_i\}\times[0,1]$ in $M$ and $\on{hol}_{\gamma_i}$ stands for the holonomy of the connection along $\gamma_i$. Notice that $f_i(\on{hol}_{\gamma_i})$ is gauge invariant, due to the $H$-invariance of $f_i$.
$$
\begin{tikzpicture}
\coordinate (sh) at (0,-1.3);
\draw[yshift=-1.3cm, gray, very thin](6,1)--(1,1)--(0,0);
\draw[gray, very thin] (1,1)--+(sh);
\draw[thick,red] (1.3,0.6)--node[right]{$\gamma_1$}+(sh) (2.3,0.2)--node[right]{$\gamma_2$}+(sh)
(3.6,0.8)--node[right]{$\gamma_3$}+(sh) (4.5,0.4)--node[right]{$\gamma_4$}+(sh);
\draw (0,0)--(5,0)--(6,1)--(1,1)--cycle;
\draw[yshift=-1.3cm] (0,0)--(5,0)--(6,1);
\draw (0,0)--+(sh) (5,0)--+(sh) (6,1)--node[right]{$\Sigma\times [0,1]$}+(sh);

\begin{scope}[yshift=-2.5cm]
\draw[fill=black!5!white] (0,0)--(5,0)--(6,1)--(1,1)--cycle;
\draw (6,0.5)node{$\Sigma$};
\draw[inner sep=1pt] (1.3,0.6)node[pnt]{}node[right]{$\,z_1$} (2.3,0.2)node[pnt]{}node[right]{$\,z_2$}
(3.6,0.8)node[pnt]{}node[right]{$\,z_3$} (4.5,0.4)node[pnt]{}node[right]{$\,z_4$};
\end{scope}

\end{tikzpicture}
$$

In particular, if $\g_0$ is a simple compact Lie algebra and $\g=\bar\g_0\oplus\g_0$, where $\bar\g_0$ stands for $\g_0$ with the opposite pairing, we can take 
$$V_+=\bar\g_0,\ V_-=\g_0,\ \h=\text{diagonal }\g_0$$
and the corresponding $\sigma$-model is then the WZW model with the group $G_0$ (not just chiral, but full). We are, however, interested in more general examples, where $V_+$ and $V_-$ are not necessarily Lie subalgebras of $\g$.
\begin{rem}
We have been somewhat vague about the groups $G$ and $H$ corresponding to the Lie algebras $\g$ and $\h$ and our spaces of fields were appropriate only for trivial $G$-bundles. To be a bit less vague, let us remark that the $V_+$-boundary condition requires a trivialization of the bundle over $\Sigma$ and the $\h$-boundary condition requires a reduction of the structure group over $\Sigma$ from $G$ to $H$. Chern-Simons theory needs a multiplicative $U(1)$-gerbe over $G$ \cite{CW}; the required data for $H$ is a trivialization of this multiplicative gerbe over $H$. This then gives rise to a gerbe over $G/H$, which is used for the global definition of the 2-dim $\sigma$-model.
\end{rem}

\subsection{Poisson-Lie T-duality}

Let us now fix $V_+\subset\g$ and choose two Lagrangian Lie subalgebras $\h,\h'\subset\g$. The above-described boundary conditions then give us two 2-dim $\sigma$-models with the targets $G/H$ and $G/H'$. These two $\sigma$-models are called \emph{Poisson-Lie T-dual} \cite{KSe} to each other. While they are not completely equivalent (the topological boundary conditions are different, one given by $\h$ and the other by $\h'$), they differ only by topological degrees of freedom.

In the case of the usual (i.e.\ abelian) T-duality, $\g$ is abelian, $G=\g/\Lambda$ is a torus ($\Lambda\subset\g$ is a lattice), and $H,H'\subset G$ are required to be tori. For a suitable choice of $\Lambda$ the seemingly different topological boundary conditions, given by $\h$ and $\h'$, actually coincide on the quantum level \cite{KSa}. This implies that the two $\sigma$-models, with the targets $G/H$ and $G/H'$, are completely equivalent. Such a statement is not (yet) known for any non-abelian $G$.\footnote{The corresponding problem of extending the Morita equivalence of non-commutative tori \cite{RS} to Poisson homogeneous spaces suggests that the condition for full equivalence is that the corresponding quantum group is trivial; see \cite{Smor}.}

\section{Diagramatics, RG flow, and the Generalized Ricci tensor}\label{sec:diag}

 In this section we shall study perturbatively the Chern-Simons theory on the ball $B^3$ with the $V_+$-boundary condition, using the propagator \eqref{gmprop}. In particular, by studying  the short-distance behaviour close to $S^2$, we shall find the 1-loop Weyl anomaly of the $V_+$-boundary condition and the resulting RG flow of the subspace $V_+\subset\g$.
 
Our motivation is the link with 2-dim $\sigma$-models and Poisson-Lie T-duality described above. Recall that this link needs Chern-Simons theory on $\Sigma\times[0,1]$, with the $V_+$-boundary condition on $\Sigma\times\{0\}$ and with a topological boundary condition on $\Sigma\times\{1\}$. However, the RG flow does not touch the bulk Chern-Simons theory and its topological boundary conditions, and being a short distance phenomenon, it is insensitive to the choice of $\Sigma$ and to the presence of other boundary components. Studying the ball model is thus sufficient for our purposes.



\subsection{Regularization}

In general, Feynman diagrams computed with the propagator \eqref{gmprop} diverge, so we need to introduce a regularization.

We choose a smooth cutoff function
$$\ell\colon S^2\to(0,\infty).$$ 
One should think of $\ell^{-2}$ as of a factor scaling the standard Riemannian metric on $S^2$. 
We use it to regularize $\omega$ (see \eqref{omega}) to
$$\omega^{(\ell)}:=\Theta_\ell\, \omega$$
where $\Theta_\ell\colon\cf_2(S^2)\to\{0,1\}$ is given by
$$
\Theta_\ell(z_1,z_2)=
\begin{cases}
0 & \text{if } d(z_1,z_2)\leq \bigl(\ell(z_1)+\ell(z_2)\bigr)/2\\
1 & \text{otherwise}
\end{cases}
$$
where $d(z_1,z_2)$ is the spherical distance of points on $S^2$.\footnote{%
A more natural choice (not using the spherical distance) is to declare that $\ell^{-2}$ is a Riemannian metric on $S^2$ in the given conformal class and that $\Theta_\ell$ is 0 iff the geodesic distance of the two points is less than 1. For our purposes these two cutoffs give the same results.}


Then we replace $P_0$ with
$$P_0^{(\ell)}=r^*\omega^{(\ell)}/2\pi i$$
and the propagator $P$ with
$$P^{(\ell)}=\bar P_0^{(\ell)}\otimes t_+ +  P_0^{(\ell)}\otimes t_-$$
 Now all the integrals converge absolutely, as explained in \cite{AS,K} using suitable compactifications of configuration spaces.
 
\begin{figure}[h]
\includegraphics{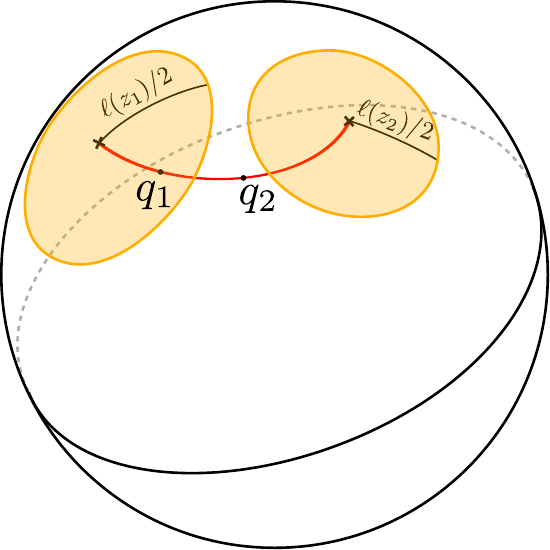}
\caption{The cutoff: the orange spherical caps should not overlap. Artist's impression.}
\end{figure}

\subsection{Diagramatics}


We can now perturbatively compute the $k$-point correlation forms
$$G_k^{(\ell)}\in\Omega^k\bigl(\cf_k(B^3)\bigr)[\![\hbar]\!]\otimes\g^{\otimes k}$$
$$G_k^{(\ell)}(x_1,\dots,x_k)=\bigl\la A(x_1)A(x_2)\dots A(x_k)\bigr\ra^{(\ell)}$$
in the usual way:

If $\Gamma$ is a tadpole-free%
\footnote{The Lie algebra factors $T_{\Gamma}$ (see below) vanish for any diagram with a tadpole, because of the 
antisymmetry of the Lie bracket.} %
graph with $k$ leaves and $n$ internal vertices, all of them 3-valent, we put $P^{(\ell)}$ on every  edge, the structure constants of $\g$
$$c\in\g^*\otimes\g^*\otimes\g^*$$
(where we identify $\g$ with $\g^*$ via the pairing on $\g$) on every vertex, contract corresponding $\g$'s with $\g^*$'s, and finally integrate over the internal vertices, i.e.\ over $\cf_n(B^3)$.  Let us denote the result by
$$\alpha^{(\ell)}_{\Gamma}\in\Omega^k\bigl(\cf_k(B^3)\bigr)\otimes\g^{\otimes k}.$$
We then have
$$G_k^{(\ell)}= 
\sum_\Gamma \frac{(i\hbar)^{k-\chi(\Gamma)}}{|\on{Aut}_0(\Gamma)|}\,\alpha^{(\ell)}_{\Gamma}$$ 
where the sum runs over all graphs with $k$ labelled leaves, $\chi(\Gamma)$ is the Euler characteristic of $\Gamma$ and $\on{Aut}_0(\Gamma)$ is the group of automorphisms of $\Gamma$ preserving the leaves.

Since the propagator is the sum of two terms
$$P^{(\ell)}=P^{(\ell)}_+ + P^{(\ell)}_-\qquad (P^{(\ell)}_+=\bar P^{(\ell)}_0\otimes t_+,\ P^{(\ell)}_-= P^{(\ell)}_0\otimes t_-),$$
it is convenient to use graphs $\Gamma'$ with signs over the edges:
$$
\begin{tikzpicture}[scale=0.6]
\coordinate (a) at (0:2cm);
\coordinate (b) at (120:2cm);
\coordinate (c) at (-120:2cm);
\coordinate (A) at (0:1cm);
\coordinate (B) at (120:1cm);
\coordinate (C) at (-120:1cm);

\draw (a)--node[proj]{\tiny $-$}(A) (b)--node[proj]{\tiny $+$}(B) (c)--node[proj]{\tiny $+$}(C);
\draw (A) to[bend right] node[proj]{\tiny $-$}(B)to[bend right] node[proj]{\tiny $+$}(C)to[bend right] node[proj]{\tiny $-$}(A);
\end{tikzpicture}
$$
Now we put $P_+^{(\ell)}$ or $P_-^{(\ell)}$ on every edge according to its sign and $c$ on every vertex. The integral over the internal vertices will give us
$$\alpha^{(\ell)}_{\Gamma'}\in\Omega^k\bigl(\cf_k(B^3)\bigr)\otimes\g^{\otimes k}$$
and for any unsigned $\Gamma$ we have $\alpha^{(\ell)}_{\Gamma}=\sum_{\Gamma'}\alpha^{(\ell)}_{\Gamma'}$, where we sum over all signed versions $\Gamma'$ of $\Gamma$.

The advantage of these signed graphs $\Gamma'$ is that we have
$$\alpha^{(\ell)}_{\Gamma'} = w^{(\ell)}_{\Gamma'}\otimes T_{\Gamma'},\qquad w^{(\ell)}_{\Gamma'}\in\Omega^k\bigl(\cf_k(B^3)\bigr),\ T_{\Gamma'}\in\g^{\otimes n}$$
where $w^{(\ell)}_{\Gamma'}$ is independent of $\g$ and of $V_+$.

Namely, $w^{(\ell)}_{\Gamma'}$ is obtained by putting $\bar P^{(\ell)}_0$ to every  $(+)$-edge, $P^{(\ell)}_0$ on every  $(-)$-edge, and integrating over the internal vertices. The Lie algebra factor $T_{\Gamma'}$ is obtained by putting $t_+$ on every $(+)$-edge, $t_-$ on every $(-)$-edge, $c$ on every vertex, and contracting corresponding $\g$'s with $\g^*$'s.

\begin{rem}[Signs] 
In the discussion above we glossed over the signs appearing in the calculation of $G_k^{(\ell)}$'s. 
We  refer the reader to \cite{AS} for the details. The idea is to see $P_\pm^{(\ell)}$ as elements of $\Omega^2(\cf_2(B^3))\otimes\bw^2(\g\oplus\g)$, which makes them symmetric under the flip of the two points, so we don't need to orient the edges. We then take the product of $P_\pm^{(\ell)}$'s over the edges to get an element of $\Omega(\cf_V(B^3))\otimes\bw(\g^V)$, where $V$ is the set of the vertices. Finally, at each internal vertex we contract with the structure constants and integrate. Since the objects/operations assigned to edges and to internal vertices are even, the outcome is well defined (i.e.\ independent of the order).

On the other hand, to define $w^{(\ell)}_{\Gamma}$ and $T_{\Gamma}$ for a signed $\Gamma$ individually (i.e.\ not just their product $\alpha^{(\ell)}_{\Gamma}$) we need additional choices on $\Gamma$, e.g.\ a cyclic order at each internal vertex, in order to fix the sign of $w^{(\ell)}_{\Gamma}$ and $T_{\Gamma}$. To make things explicit, we shall always use the anticlockwise order in the diagrams that we draw.
\end{rem}


\subsection{Generalized Ricci tensor and conformal invariance}
We shall say that a signed diagram $\Gamma$ is convergent if the limit
$$\lim_{\epsilon\to 0_+}w_{\Gamma}^{(\epsilon\ell)}$$
exists and is independent of $\ell$. 

We shall see in a moment that the only divergent 1-loop 1PI diagram is 
\begin{equation}\label{eyediagram}
D:=
\begin{tikzpicture}[scale=0.6, baseline=-0.6ex]
\coordinate (u) at (-0.5,0);
\coordinate (x) at (1,0);
\coordinate (y) at (3,0) {};
\coordinate (v) at (4.5,0);
\node (plus) at (2,0.8) [proj] {\tiny $+$};
\node (minus) at (2,-0.8) [proj] {\tiny$-$};
\draw (u)--node[proj]{\tiny $+$}(x) (y)--node[proj]{\tiny $-$}(v);
\draw (x) to[bend left] (plus) (plus)to[bend left] (y)
      (x) to[bend right] (minus) (minus)to[bend right] (y) ;
\end{tikzpicture}
\end{equation}
As a result, the 1-loop condition for conformal invariance of the $V_+$-boundary condition is 
$$T_{D}=0$$
(note that $T_{D} = 0$ in the case of the chiral boundary condition, where $V_{+} = 0$).

The element $-T_{D}\in V_+\otimes V_-$ is known as the \emph{generalized Ricci tensor} of the generalized metric $V_+\subset\g$ \cite{CSW,G,JV,SV}. The 1-loop conformality condition is thus the vanishing of the generalized Ricci tensor. 

\begin{rem}
The 1-loop conformality condition $T_D=0$ (and the corresponding role of $T_D$ as a $\beta$-function, see \S\ref{ss:RGflow}) was discovered in \cite{SST} for a 2-dim model which can be seen as a Hamiltonian version of the Chern-Simons model with the $V_+$-boundary condition. This somewhat miraculous result of \emph{op.cit.} thus gets a natural explanation, coming from a single Chern-Simons diagram.
\end{rem}

\subsection{Effective action and Weyl anomaly}
As usual, it is more economical to study the  1PI effective action $S^{(\ell)}_\text{eff}$ rather than the correlation forms $G_k^{(\ell)}$. It is a functional on our BV space of fields
$$\Omega_{\bar b}(B^3,V_+)[1] \oplus \Omega_{ b}(B^3,V_-)[1] \subset \Omega(B^3,\g)[1] $$
given by
$$S^{(\ell)}_\text{eff}(A)= S(A) + 
\sum_\Gamma \frac{(i\hbar)^{1-\chi(\Gamma)}}{|\on{Aut}(\Gamma)|} S_\Gamma^{(\ell)}(A).$$
Here $\Gamma$ runs over all 1PI graphs with at least one loop, $1-\chi(\Gamma)$ is the number of loops of $\Gamma$, and $\on{Aut}(\Gamma)$ is the group of all automorphisms of $\Gamma$. Finally, $S_\Gamma^{(\ell)}(A)$ is obtained by putting $A$ on every leaf, $P^{(\ell)}$ on every internal edge, $c$ on every internal vertex, contracting corresponding $\g$'s and $\g^*$'s, and integrating. 


Let us now study the 1-loop part of $S_\text{eff}$. Only the term quadratic in $A$ is divergent (see Appendix). It is given by the diagram
$$
\begin{tikzpicture}[baseline=-1,scale=0.8]
\coordinate (a) at (0,0);
\coordinate (b) at (2,0);
\node[left] (A1) at (-0.2,0) {${\displaystyle\frac {i\hbar} 4}\quad A$};
\node[right] (A2) at (2.2,0) {$A$};

\draw (a) to[out=60,in=120] (b) to[out=-120,in=-60] (a);
\draw[gray] (a)--(A1);
\draw[gray] (b)--(A2);

\end{tikzpicture}
$$
Since $(P_0^{(\ell)})^2=(\bar P_0^{(\ell)})^2=0$, the diagram is equal to
$$
\begin{tikzpicture}[baseline=-1, scale=0.8]
\coordinate (a) at (0,0);
\coordinate (b) at (2,0);
\node[left] (A1) at (-0.2,0) {${\displaystyle\frac {i\hbar} 2}\quad A$};
\node[right] (A2) at (2.2,0) {$A$};

\draw (a) to[out=60,in=120]node[proj]{\tiny$+$} (b) to[out=-120,in=-60]node[proj]{\tiny$-$} (a);
\draw[gray] (a)--(A1);
\draw[gray] (b)--(A2);

\end{tikzpicture}
$$
To get its divergent part, we use the following lemma which can be proved by a straightforward calculation (see Appendix).
\begin{lem}\label{lem:CS}
If $\ell_1,\ell_2\colon S^2\to(0,\infty)$ are smooth, $\alpha,\beta\in\Omega(B^3)$, and if $p_1$ and $p_2$ are the projections $\cf_2(B^3)\to B^3$, then
\begin{multline*}
\lim_{\epsilon\to0_+}\int_{\cf_2(B^3)}r^*(\Theta_{\epsilon\ell_1}-\Theta_{\epsilon\ell_2})\;P_0\bar P_0\;p_1^*\alpha\; p_2^*\beta\\
=-\frac1{2\pi} \int_{S^2}\log(\ell_1/\ell_2)\;(\alpha^{(1)}|_{S^2})\,*(\beta^{(1)}|_{S^2}) 
\end{multline*}
where $\alpha^{(1)}$ is the 1-form part of $\alpha$ and $*$ is the Hodge $*$ on $S^2$.
\end{lem}
We thus get the 1-loop Weyl anomaly
\begin{multline}\label{conf-anom}
\lim_{\epsilon\to0_+}\Bigl( S_\text{eff}^{(\epsilon\ell_1),\text{1-loop}}(A) - S_\text{eff}^{(\epsilon\ell_2),\text{1-loop}}(A)\Bigr)\\
=-\frac{\hbar}{2\pi}\int_{S^2} \log(\ell_1/\ell_2)\;\la T_{D}, A^{(1)}_+ A^{(1)}_-\ra
\end{multline}
where $A_\pm$ is the component of $A$ with the values in $V_\pm$. Notice that on $S^2$ we have $A^{(1)}_+A^{(1)}_+=A^{(1)}_-A^{(1)}_-=0$ because of the boundary condition, which explains why only the diagram $D$ appears in the result.

\subsection{RG flow of $V_+$}\label{ss:RGflow}
\subsubsection{Renormalization}
If the generalized Ricci tensor $-T_{D}\in V_+\otimes V_-$ is non-zero, the limit $\lim_{\epsilon\to0_+}S_\text{eff}^{(\epsilon\ell),\text{1-loop}}$ doesn't exist.
From \eqref{conf-anom} we get
\begin{equation}\label{Seff-asymp}
  \lim_{\epsilon\to0_+} \epsilon\,\frac d{d\epsilon}\; S_\text{eff}^{(\epsilon\ell),\text{1-loop}}(A)= -\frac\hbar{2\pi}\int_{S^2} \la T_{D}, A^{(1)}_+ A^{(1)}_-\ra.  
\end{equation}

To get a meaningful effective action in the limit $\epsilon\to0$ we have to renormalize $V_+$, i.e.\ make $V_+(\epsilon)\subset\g$  a function of $\epsilon$ such that
$$
\lim_{\epsilon\to0_+}S_\text{eff}^{(\epsilon\ell),\text{1-loop}}(A,V_+(\epsilon))
$$
exists (where we made the dependence of $S_\text{eff}$ on $V_+$ explicit). More precisely, we impose
\begin{equation}\label{RG-howto}
\lim_{\epsilon\to0_+} \epsilon\,\frac d{d\epsilon}\; S_\text{eff}^{(\epsilon\ell),\text{1-loop}}(A,V_+(\epsilon))=0
\end{equation}
and from that and from \eqref{Seff-asymp} get an ODE for $V_+$, i.e.\ its RG flow.

\subsubsection{Field redefinition}
This usual approach has to be slightly modified, however, because the space of fields depends on $\epsilon$ (as $V_+$, and thus the boundary condition, depends on $\epsilon$). We will redefine the fields so that the boundary condition doesn't change, and instead get a change of the action $S$.

If we have an infinitesimal $\delta R\in\bw^2\g=\mf{so}(\g)$, let us deform $V_+$ to $(1+\delta R)V_+$ (if $v\in\g$, by $\delta R\,v\in\g$ we mean the element given by $\la\delta R\,v,w\ra=\la \delta R,v\otimes w\ra$ for all $w\in\g$). We can then redefine the fields so that they satisfy the deformed boundary condition via 
$$A\to A+\delta A, \quad\delta A= f\,\delta R\,A,$$
where $f\in C^\infty(B^3)$ is an arbitrary function s.t.\ $f|_{S^2}=1$ (if needed, we can choose $f$ so that $\delta A$ in non-zero only close to the boundary).

The variation of the (classical) action that we obtain is
$$S(A+\delta A)-S(A) = 
\int_{B^3}\bigl\la f\,\delta R, A\,dA+\tfrac12A[A,A]\bigr\ra + \frac12\int_{S^2}\la \delta R,AA\ra.
$$
 Let us note that the first integral is BV-exact, generated by
 $$\frac12\int_{B^3}\la f\,\delta R, AA\ra.$$
In other words, the change of the boundary condition is equivalent (up to a BV-exact term) to not changing the boundary condition, and adding the boundary term
$$\frac12\int_{S^2}\la \delta R,AA\ra$$
to the action functional instead.

\subsubsection{The RG flow}
Now we can interpret \eqref{RG-howto}: if 
\begin{subequations}\label{RG}
\begin{equation}
V_+(\epsilon+d\epsilon)=\bigl(1+\frac{d\epsilon}\epsilon\,\hbar\, B\bigr)V_+(\epsilon)
\end{equation}
 for a suitable $B\in\bw^2\g$, then \eqref{RG-howto} gives us (in the limit $\epsilon\to0$)
$$-\frac\hbar{2\pi}\int_{S^2} \la T_{D}, A^{(1)}_+ A^{(1)}_-\ra +\frac\hbar2\int_{S^2}\la B,AA\ra=0$$
and so we can set
\begin{equation}
B=\frac1{2\pi} (T_{D}^{\vphantom{o}}-T_{D}^{\hphantom{x}\text{op}}).
\end{equation}
We thus found the 1-loop RG flow of $V_+$ in our model, with $\hbar B$ playing the role of the $\beta$-function. 
\end{subequations}

\section{Further developments}
In this section we shall very briefly summarize natural generalizations of what we did in this paper; they will be treated in more detail elsewhere. Before starting, let us mention that the model from \S\ref{sec:diag} should also be developed in more detail, whether it is the calculation of the Weyl anomaly \eqref{conf-anom} to higher powers of $\hbar$, or a calculation of a generalized Knizhnik-Zamolodchikov connection for  Wilson lines with endpoints on $S^2$.

\subsection{Courant $\sigma$-model and generalized Ricci tensor}
Chern-Simons theory with the $V_+$-boundary condition 
has a natural generalization to a Courant $\sigma$-model (again a 3-dim TFT) with a boundary condition given, again, by a generalized metric in the corresponding Courant algebroid. Such a model is needed for a formulation of the Poisson-Lie T-duality with spectators \cite{PSV,S}. Moreover, if the Courant algebroid is exact, this model is directly equivalent to a 2-dim $\sigma$-model.

\subsubsection{Courant $\sigma$-model}
Let us describe Courant $\sigma$-models supposing that the underlying Courant algebroid is trivial as a vector bundle and that its base is a vector space (this trivialization is a part of our gauge fixing); we refer the reader to \cite{R} for an invariant and global description. 

We need a vector space $W$ (the base of the Courant algebroid) and another vector space $V$ with an (indefinite) inner product (the fibre of the Courant algebroid). Finally, we need a function $C$ of degree 3 on the graded symplectic manifold 
$$W\oplus V[1]\oplus W^*[2]$$
satisfying the classical master equation $\{C,C\}=0$. Written in components, $C$ is of the form
$$C(x,A,p)=\frac16\,c_{abc}(x)\,A^aA^bA^c + \rho^i_a(x)\,p_iA^a.$$

If now $M$ is a closed oriented 3-manifold, the space of fields of the Courant $\sigma$-model on $M$ is
$$\Omega(M,W)\oplus\Omega(M,V)[1]\oplus\Omega(M,W^*)[2]$$
(with the 3 components still denoted by $x$, $A$, and $p$)
and the action functional is
$$S(x,A,p)=\int_M\la p,dx\ra + \frac12 \la A,d A\ra + C(x,A,p).$$

\subsubsection{Boundary condition}
If $M$ has a boundary $\Sigma$, our boundary condition is given as follows. We choose an orthogonal splitting
$$V=V_+\oplus V_-$$
such that the pairing is positive definite on $V_+$ and negative definite on $V_-$ (i.e.\ a generalized metric), and impose that the boundary fields are in the subcomplex
\begin{multline*}
\Omega(\Sigma,W)\oplus\bigl(\Omega_{(+)}(\Sigma,V_+)\oplus\Omega_{(-)}(\Sigma,V_-)\bigr)[1]\\
\subset\ \Omega(\Sigma,W)\oplus\Omega(\Sigma,V)[1]\oplus\Omega(\Sigma,W^*)[2]. 
\end{multline*}
The action functional remains the same.

\subsubsection{Propagators}
Let us now suppose that $M=H^3\subset\R^3$ is the upper half space (we do not take $M=B^3$ since our complex of fields would have a non-trivial cohomology, and for our purposes $M=H^3$ is sufficient). The total propagator is the sum
$$P=\bar P_0\otimes t_+ +  P_0\otimes t_- + P_1\otimes\id + P_1^{op}\otimes\id^{op}.$$
Here $P_0$ is as before, $t_\pm\in S^2V_\pm$ as well, 
$$P_1\in\Omega^2(\cf_2(H^3))$$ is a suitable closed 2-form vanishing when the first point is on the boundary, and $\id\in W^*\otimes W$ is the identity $W\to W$. Graphically we shall represent the terms in this sum as
$$P=\ 
\begin{tikzpicture}[baseline=-0.1cm]
\draw (0,0)--node[proj]{\tiny$+$}(1,0);
\end{tikzpicture}
\ +\ 
\begin{tikzpicture}[baseline=-0.1cm]
\draw (0,0)--node[proj]{\tiny$-$}(1,0);
\end{tikzpicture}
\ +\ 
\begin{tikzpicture}[baseline=-0.1cm]
\draw[dotted,->-=0.65] (0,0)--(1,0);
\end{tikzpicture}
\ +\ 
\begin{tikzpicture}[baseline=-0.1cm]
\draw[dotted,->-=0.65] (1,0)--(0,0);
\end{tikzpicture}
$$

For $P_1$ we shall take, following \cite{K2},
$$P_1=r_1^*\omega_{S^2}/4\pi$$
where $\omega_{S^2}$ is the area form on $S^2$ and $r_1\colon\cf_2(H^3)\to S^2$ sends a pair of points $(q_1,q_2)$ to the unit vector tangent at $q_1$ to the hyperbolic geodesic passing through $q_1$ and $q_2$.

\subsubsection{Diagramatics}
Let us pick a point $x_0\in W$ around which we shall do the perturbative expansion. We have two types of vertices,
$$
\begin{tikzpicture}[scale=0.6]
\draw(0,0)--(1,0)(0,0)--(-0.5,0.866)(0,0)--(-0.5,-0.866);
\node at (2,0){and};
\draw (3,0)--(4.5,0)node[anch]{};
\draw[dotted,->-=0.65] (4.5,0)--(6,0);
\end{tikzpicture}
$$
At the first vertex we put $c_{abc}(x_0)$ and at the second one $\rho^i_a(x_0)$. Furthermore, each of these vertices can have an arbitrary number of incoming dotted lines; we need to differentiate $c_{abc}$'s and $\rho^i_a$'s for each such line.
Finally, the solid lines should be split to the sum of \tikz[baseline]{\draw (0,0.5ex)--node[proj]{\tiny$+$}(1,0.5ex);} and  \tikz[baseline]{\draw (0,0.5ex)--node[proj]{\tiny$-$}(1,0.5ex);}.

Now each diagram $\Gamma$ evaluates as the product of two factors $w_\Gamma$ and $T_\Gamma$: $w_\Gamma$ is an integral of a product of $P_0$'s, $\bar P_0$'s, and $P_1$'s, and $T_\Gamma$ a contraction of $c(x_0)$'s, $\rho(x_0)$'s, and of their derivatives.

\subsubsection{1-loop RG flow and Generalized Ricci tensor}
We need to regularize $P_0$ to $P_0^{(\ell)}$ as above, though $P_1$ doesn't need to be regularized. A simple calculation (see Appendix) shows that the divergent part of the 1-loop effective action is given by $T_{D'}$ (a replacement of $T_D$) where
\begin{equation}\label{ggric}
D'=
\begin{tikzpicture}[scale=0.6, baseline=-0.6ex]
\node (u) at (-0.5,0) {};
\coordinate (x) at (1,0);
\coordinate (y) at (3,0) {};
\node (v) at (4.5,0) {};
\node (plus) at (2,0.8) [proj] {\tiny $+$};
\node (minus) at (2,-0.8) [proj] {\tiny$-$};
\draw (u)--node[proj]{\tiny $+$}(x) (y)--node[proj]{\tiny $-$}(v);
\draw (x) to[bend left] (plus) (plus)to[bend left] (y)
      (x) to[bend right] (minus) (minus)to[bend right] (y) ;
\end{tikzpicture}
+\;\frac12
\begin{tikzpicture}[scale=0.6, baseline=-0.6ex]
\node (u) at (-1.5,-1) {};
\node (v) at (1.5,-1) {};
\coordinate (x) at (0,0);
\node (minus) at (0,1) [proj] {\tiny$+$};
\draw (u)--node[proj]{\tiny $+$}(x)--node[proj]{\tiny $-$}(v) (x)to[out=75, in=-75](minus);
\node (rho) at (-0.8,0.7) [anch]  {};
\draw (minus) to[out=140,in=90] (rho);
\draw[dotted,->-=0.65] (rho) to[out=-70,in=160] (x);
\end{tikzpicture}
-\;\frac12
\begin{tikzpicture}[scale=0.6, baseline=-0.6ex]
\node (u) at (-1.5,-1) {};
\node (v) at (1.5,-1) {};
\coordinate (x) at (0,0);
\node (minus) at (0,1) [proj] {\tiny$-$};
\draw (u)--node[proj]{\tiny $+$}(x)--node[proj]{\tiny $-$}(v) (x)to[out=75, in=-75](minus);
\node (rho) at (-0.8,0.7) [anch]  {};
\draw (minus) to[out=140,in=90] (rho);
\draw[dotted,->-=0.65] (rho) to[out=-70,in=160] (x);
\end{tikzpicture}
\end{equation}

This expression is, up to the sign, the \emph{generalized Ricci tensor} of the generalized metric $V_+$ in our Courant algebroid. As a result we see that the generalized Ricci tensor gives us the 1-loop RG flow.

\begin{rem}
Let us be more precise about the appearance of $T_{D'}$: we have
\begin{multline*}
\lim_{\epsilon\to0_+}\Bigl( S_\text{eff}^{(\epsilon\ell_1),\text{1-loop}}(x,A,p) - S_\text{eff}^{(\epsilon\ell_2),\text{1-loop}}(x,A,p)\Bigr)\\
=-\frac{\hbar}{2\pi}\int_{S^2} \log(\ell_1/\ell_2)\;\la T_{D'}(x), A^{(1)}_+ A^{(1)}_-\ra  
\end{multline*}
where $T_{D'}(x)$ denotes $T_{D'}$ computed with $c$'s and $\rho$'s at the point $x\in W$.  
\end{rem}

This problem requires, however, a more detailed analysis. The generalized Ricci tensor depends on the gauge fixing (though only up to a redefinition of fields) -- in fact, it depends on a so-called divergence operator \cite{G}, which seems closely related to the tadpole diagram
$$
\begin{tikzpicture}
\clip(-0.1,-0.1) rectangle (2,1);
\node[anch] (a) at (1,0){};
\draw (0,0)--(a);
\draw[dotted,->-=0.9] (a)..controls (2.5,0) and (1,1.5)..(a);
\end{tikzpicture}
$$
in our model. Moreover, a conformal invariance should also require a kind of a dilaton field (indeed, when the underlying Courant algebroid is exact, our model is equivalent to a 2-dim $\sigma$-model with the target $W$, and in that case we should certainly see a dilaton field). The results of \cite{SV2} suggest that this generalized dilaton field should be a half-density and the conformality condition should require it to be an eigenvector of a natural Laplace operator. We shall leave these problems to a future work.

\subsection{AKSZ models, higher dualities, and RG flow}
The link between the Poisson-Lie T-duality of 2-dim $\sigma$-models and the 3-dim Chern-Simons (or Courant) theory has a natural generalization to higher dimensions, when (possibly higher gauge) $n$-dim models are obtained from boundary conditions of $n+1$-dim topological models, namely of $n+1$-dim AKSZ models \cite{AKSZ}. We refer the reader to \cite{PSV} for details.

The AKSZ model is constructed out of a dg symplectic manifold $X$ with symplectic form of degree $n$ ($X=\g[1]$ in the case of Chern-Simons theory). If $M$ is a $n+1$-dim manifold with a boundary $\Sigma$, a boundary condition for the AKSZ model is an exact Lagrangian dg submanifold $\mc F$ in the symplectic space of all boundary fields
$$\mc X:=\on{Maps}(T[1]\Sigma,X).$$

While the AKSZ model is topological, the boundary condition $\mc F$ might be not (as was the case of the $V_+$-boundary condition). In particular, under perturbative quantization, $\mc F$ might become cutoff-dependent, i.e.\ at least naively we should expect a vector field (the RG flow) on the space (the exact Lagrangian Grassmannian) $\mc{G}_\mc X$ of all $\mc F$'s.

Let us only describe how to calculate the RG flow.
Since for any $\mc F\in\mc{G}_\mc X$ we have
$$T_\mc F \mc{G}_\mc X =C^\infty(\mc F), $$
the RG flow should assign a function on $\mc F$, i.e.\ a functional on the space of the boundary fields, to every $\mc F$. This functional is, following the calculation from \S\ref{ss:RGflow}, equal to
\begin{equation}\label{akszrg}
-\lim_{\epsilon\to0_+} \epsilon\,\frac d{d\epsilon}\; S_\text{eff}^{(\epsilon\ell)}.
\end{equation}

\appendix

\section{1-loop divergent diagrams}
\subsection{Chern-Simons}
Let us prove Lemma \ref{lem:CS}. In place of the ball $B^3$ we shall be using the closed upper half space $H^3$ (with the boundary identified with $\C$), and we shall suppose that $\alpha$ and $\beta$ are compactly supported. We shall use the following coordinates on $\cf_2(H^3)$: if $(q_1,q_2)\in\cf_2(H^3)$, let $(z,z+u):=r(q_1,q_2)\in\C^2$, and let the vertical projections of $q_1$ and $q_2$ to the plane be $z+t_1u$ and $z+t_2u$, $0\leq t_1<t_2\leq1$. We want to compute
$$I:=
\lim_{\epsilon\to0_+}\int_{\cf_2(H^3)}r^*(\Theta_{\epsilon\ell_1}-\Theta_{\epsilon\ell_2})\;\frac1{4\pi^2}\,\frac{dz\,du\,d\bar z\,d\bar u}{|u|^4}\,p_1^*\alpha\, p_2^*\beta
$$

Supposing that $\alpha$ and $\beta$ are 1-forms, let us write $\alpha$ as $\alpha=a_+dz + a_-d\bar z + a_h dh$, where $h$ is the height coordinate on $H^3$, and similarly 
$\beta=b_+dz + b_-d\bar z + b_h dh$. The terms with $dh$ do not contribute to the limit, and so we can suppose that they vanish. We then have
\begin{align*}
p_1^*\alpha&=a_+\, d(z+t_1 u) + a_-\, d(\bar z + t_1\bar u)\\
p_2^*\beta&=b_+\, d(z+t_2 u) + b_-\, d(\bar z + t_2\bar u)
\end{align*}
and so
\begin{multline*}
I=-\frac1{4\pi^2}\lim_{\epsilon\to0_+}\int_{\cf_2(H^3)}r^*(\Theta_{\epsilon\ell_1}-\Theta_{\epsilon\ell_2})\;\frac{du\,d\bar u}{|u|^2}\,dz\,d\bar z\,dt_1\,dt_2\,(a_+b_-+a_-b_+) \\
=\frac i{4\pi^2}\lim_{\epsilon\to0_+}\int_{\cf_2(H^3)}r^*(\Theta_{\epsilon\ell_1}-\Theta_{\epsilon\ell_2})\;\frac{du\,d\bar u}{|u|^2}\,dt_1\,dt_2\,\alpha *\beta.
\end{multline*}
The integral of 
$$r^*(\Theta_{\epsilon\ell_1}-\Theta_{\epsilon\ell_2})\;\frac{du\,d\bar u}{|u|^2}$$
gives $-4\pi i\log(\ell_1/\ell_2)$, the integral of $dt_1\,dt_2$ gives $1/2$, and the orientation changes the sign, so in the end
$$I=-\frac1{2\pi}\int_\C\log(\ell_1/\ell_2)\,\alpha*\beta$$
as we wanted to show.

If one of $\alpha,\beta$ is a 0-form and the other a 2-form, a similar calculation gives $I=0$.

\subsection{Courant model}
We have two new divergent 1-loop diagrams, namely
$$
\begin{tikzpicture}[scale=0.8, baseline=1ex]
\coordinate (x) at (0,0);
\node (minus) at (0,1) [proj] {\tiny$+$};
\draw (-0.3,-0.3)--(x)--(0.3,-0.3) (x) to[out=75, in=-75](minus);
\node (rho) at (-0.8,0.7) [anch]  {};
\draw (minus) to[out=140,in=90] (rho);
\draw[dotted,->-=0.65] (rho) to[out=-70,in=160] (x);
\end{tikzpicture}
\qquad\text{and}\qquad
\begin{tikzpicture}[scale=0.8, baseline=1ex]
\coordinate (x) at (0,0);
\node (minus) at (0,1) [proj] {\tiny$-$};
\draw (-0.3,-0.3)--(x)--(0.3,-0.3) (x) to[out=75, in=-75](minus);
\node (rho) at (-0.8,0.7) [anch]  {};
\draw (minus) to[out=140,in=90] (rho);
\draw[dotted,->-=0.65] (rho) to[out=-70,in=160] (x);
\end{tikzpicture}
$$
Let us compute the divergent part of the first one (the second one is similar).

Given $\alpha\in\Omega^2(\C)$ with a compact support, we want to find
$$J:=
\lim_{\epsilon\to0_+}\int_{\cf_2(H^3)}r^*(\Theta_{\epsilon\ell_1}-\Theta_{\epsilon\ell_2})\;P_0P_1\,p_2^*\alpha.
$$
Using the same coordinates as above, we have
$$P_1=\frac1{4\pi i}(\frac{du}u - \frac{d\bar u}{\bar u})\,dt_1$$
so
$$
J=\frac1{8\pi^2}\lim_{\epsilon\to0_+}\int_{\cf_2(H^3)}r^*(\Theta_{\epsilon\ell_1}-\Theta_{\epsilon\ell_2})
\frac{dz\,du}{u^2}\,\frac{d\bar u}{\bar u}\,dt_1\,p_2^*\alpha
$$
Writing $\alpha$ as $\alpha=f\,dz\,d\bar z$ (we can again ignore the vertical component of $\alpha$) we then get
\begin{align*}
J&=\frac1{8\pi^2}\lim_{\epsilon\to0_+}\int_{\cf_2(H^3)}r^*(\Theta_{\epsilon\ell_1}-\Theta_{\epsilon\ell_2})
\frac{dz\,du}{u^2}\,\frac{d\bar u}{\bar u}\,dt_1\,f\,u\,dt_2\,d\bar z\\
&=\frac1{8\pi^2}\lim_{\epsilon\to0_+}\int_{\cf_2(H^3)}r^*(\Theta_{\epsilon\ell_1}-\Theta_{\epsilon\ell_2})
\frac{du\,d\bar u}{|u|^2}\,dt_1\,dt_2\,\alpha=-\frac1{4\pi i} \int_\C\log(\ell_1/\ell_2)\,\alpha.
\end{align*}
The prefactor $1/4\pi$, as compared to the prefactor $1/2\pi$ in $I$, is responsible for the coefficients $1/2$ in  \eqref{ggric}.

\subsection{Diagrams with at least 3 vertices converge}
Let us now consider a general 1-loop diagram of length $n$ ($n=5$ on the picture)
$$
\begin{tikzpicture}
\draw (0:1cm)node[anch,label=0:$\alpha_1$]{}--(72:1cm)node[anch,label=72:$\alpha_2$]{}--(144:1cm)node[anch,label=144:$\alpha_3$]{}--(-144:1cm)node[anch,label=-144:$\alpha_4$]{}--(-72:1cm)node[anch,label=-72:$\alpha_5$]{}--cycle;
\end{tikzpicture}
$$
where on each edge we put one of the 2-forms $P_0$, $\bar P_0$, $P_1$, or $P_1^{op}$,  on each vertex a compactly supported form $\alpha_i\in\Omega(H^3)$, and finally integrate over $\cf_n(H^3)$. Notice that the total degree of $\alpha_i$'s must be $n$, as we need to integrate a $3n$-form.

Let us first remove a tubular neighbourhood of the diagonal $\C\subset H^3\subset_\text{diag}(H^3)^n$ from $\cf_n(H^3)$. To be specific, we fix $\epsilon>0$ and remove the configurations where all the $n$ points $q_1,\dots,q_n\in H^3$ satisfy $d(q_i,q_0)<\epsilon$ for some $q_0\in\C$ (here $d$ is the Euclidean distance). Then the integral, which we will denote by $I_\epsilon$, converges (because there are ``no divergent subdiagrams'' of our diagram).

Now we want to see that $\lim_{\epsilon\to0}I_\epsilon$ exists if $n\geq3$. For that we use the $\epsilon\to0$ behaviour of $\epsilon\frac d{d\epsilon}I_\epsilon$:
$$\epsilon\frac d{d\epsilon}I_\epsilon=O(\epsilon^{n-2}).$$
This behaviour follows from a simple scaling argument: the propagators are scale invariant, and if we locally approximate $\alpha_i$'s by constant forms, they scale with the $n$-th power (since their total degree is $n$; the exponent shift $-2$ in $\epsilon^{n-2}$ comes from scaling along $\C$). From this it follows that if $n\geq3$ then $\lim_{\epsilon\to0}I_\epsilon$ indeed exists.

\end{document}